\documentclass[a4paper,11pt]{article}
\usepackage{pos}

\usepackage{bm}
\usepackage{epsfig}
\usepackage{graphicx}
\usepackage{amsmath}

\usepackage{amssymb}
\usepackage{mathrsfs}
\usepackage{color}
\usepackage[usenames,dvipsnames]{xcolor}
\usepackage{hyperref}
\usepackage{feynmp-auto}
\usepackage{tikz}
\usepackage{braket}
\usepackage[caption=false]{subfig}
\usepackage{multirow}
\usepackage{array}
\usepackage{slashed}
\usepackage{rotating}
\usepackage{caption}
\usepackage{subcaption}
\usepackage{bbold}
\usepackage{xcolor}

\usepackage[normalem]{ulem}
\usepackage{marginnote}

\newcommand\identity{1\kern-0.25em\text{l}}
\newcommand\vb{\text{v}}

\newcommand\bv{{\bf b}}

\newcommand\Pc{{\mathcal P}}

\newcommand{\Otsooct}{\braket{\mathcal{O}^{V}(^3S_1^{[8]})}}
\newcommand{\Oosz}{\braket{\mathcal{O}^{V}(^1S_0^{[8]})}}
\newcommand{\Otpz}{\braket{\mathcal{O}^{V}(^3P_J^{[8]})}}

\title{Factorizing quarkonium LDMEs and TMDSTFs using effective field theory}

\author*[a]{Marston Copeland}

\affiliation[a]{Theoretical Division, MS. B283, Los Alamos National Laboratory,\\  Los Alamos, NM 87545, USA}

\emailAdd{pmcopeland@lanl.gov}

\abstract{We use effective field theory to factorize production matrix elements that appear in the NRQCD framework for quarkonium cross sections. By applying a Hubbard–Stratonovich transformation and appropriate field redefinitions, we show that the soft and ultrasoft sectors of NRQCD can be decoupled from the heavy quark and antiquark fields in a hybrid vNRQCD/pNRQCD Lagrangian at leading order in the velocity power-counting. This enables us to re-factorize quarkonium production matrix elements in terms of matrix elements of color-singlet composite fields, which we can write as the wave-function at the origin, and state independent vacuum correlators of chromo-electric and chromo-magnetic gluon fields. This approach verifies powerful relationships between the LDMEs of different S-wave quarkonia originally derived using pNRQCD. Additionally, it allows us to derive new relationships for the production matrix elements used in the transverse momentum dependent factorization (TMD) framework, known as TMD soft transition functions, providing a much stronger set of constraints on these nonperturbative operators. This work significantly advances our understanding of quarkonium production, particularly in the TMD framework.}

\FullConference{The 33rd International Workshop on Deep Inelastic Scattering and Related Subjects (DIS2026)\\
4 - 8 May 2026\\
Bologna, Italy\\}

\begin{document}

\noindent\makebox[\dimexpr\linewidth-1cm\relax][r]{LA-UR-26-24750}

\maketitle

\section{Introduction}

Recent innovations in quarkonium production theory have shown that soft and ultrasoft dynamics can play a critical role, meaning that it is necessary to go beyond the traditional NRQCD framework for production, and instead use either the velocity NRQCD (vNRQCD) or potential NRQCD (pNRQCD) formalisms. For example, in the TMD factorization framework, when a final state quarkonium is produced with $P_T \sim m\vb$, the transverse momentum created during the production process stems almost entirely from soft gluon radiation. This necessitates new $P_T$ dependent production operators to replace the long-distance matrix elements (LDMEs) of NRQCD. These are known as TMD soft transition functions (TMDSTFs) \cite{Copeland:2025vop} and TMD shape functions (TMDShFs) \cite{ Echevarria:2024idp, Echevarria:2025oab} which are defined explicitly using vNRQCD \cite{Copeland:2025vop,Echevarria:2024idp} .
In the collinear factorization framework, matching the LDMEs onto the composite fields of pNRQCD shows that the matrix elements can be written in terms of universal vacuum correlators of soft chromo-magnetic and chromo-electric field operators and the quarkonium's wave function at the origin \cite{Brambilla:2022ayc}. This unorthodox result, if true, leads to strong constraints on the LDMEs of different S-wave vector quarkonium states, reducing the number of free parameters in NRQCD production cross sections from 12 to 3 \cite{Brambilla:2022ayc}. 
%
%
However, upon a naive inspection, it is difficult to see how these pNRQCD results can be reproduced using the vNRQCD approach, which is concerning given that these two frameworks are supposedly equivalent. In this proceeding, we summarize recent findings from a novel analysis of quarkonium LDMEs and TMDSTFs in the vNRQCD framework \cite{Copeland:2026yqa}. We discuss how to factorize soft fields from the ultrasoft in quarkonium production matrix elements using a Hubbard-Stratonovich transformation on vNRQCD. We find that the soft-sector can only be decoupled in production matrix elements due to a unique feature of the calculation - the quark/antiquark fields are produced at a point. We summarize the factorized results and present new subleading P-wave operator contributions. Lastly we discuss how this work can be applied to the TMD framework and derive important new constraints on the TMDSTFs. 

\section{Hubbard-Stratonovich Transformations for Quarkonium Production Matrix Elements}


Two decades ago, Ref.~\cite{Fleming:2005pd} showed that a Hubbard Stratonovich transformation can be used to relate the leading order Lagrangians between the pNRQCD and vNRQCD formalisms, hinting at their equivalence. The Hubbard-Stratonovich transformation is an exact mathematical operation which replaces four fermion operators in a Lagrangian with auxilary bosonic fields coupled to fermions. In the vNRQCD framwork, this can be used to replace the non-local potential operator with color-singlet and color-octet composite fields coupled to quark-antiquark pairs. After integrating out the fermions completely, this transformation exactly reproduces the leading pNRQCD Lagrangian, whereas integrating out the composite fields returns the theory to the original vNRQCD formalism. Therefore, this method serves as an interpolation between the two frameworks. 

One important caveat prevents a strict equivalence between the vNRQCD and pNRQCD frameworks. In the vNRQCD formalism, there are explicit interactions between the quarks and gluons with soft scaling. These prevent a total factorization between the quark/antiquark and soft sectors, which contradicts with the pNRQCD philosophy. Ref.~\cite{Copeland:2026yqa} showed that these soft interactions lead to new operators during the Hubbard-Stratonovich transformation, and can play a role when there are soft final states in the system. However, when the radius of the $Q\bar{Q}$ goes to zero, as it does in quarkonium  production operators, these contributions can be shown to vanish \cite{Copeland:2026yqa} and hence a factorization between soft and ultrasoft is permitted. These proceedings summarize how these effective field theory techniques can be used to factorize the quarkonium production operators in vNRQCD \cite{Copeland:2025vop}. For a detailed analysis of this work, see Ref.~\cite{Copeland:2026yqa}.  


\subsection{S-wave matrix elements}
\begin{figure}
\centering
\vspace{-.7cm}
\includegraphics[width = .8\linewidth]{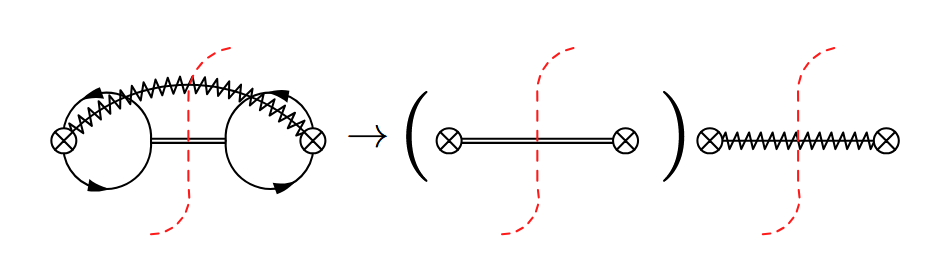}
\vspace{-.3cm}
    \caption{Factorizing out the soft chromo-magnetic dipole contributions in the $^1S_0^{[8]}$ channel. Zigzag with a line through the center represents the chromo-magnetic field. Double lines represent the color-singlet composite field.}
    \vspace{-.2cm}
    \label{fig: 1S08 quark loops}
\end{figure}

We begin with the color-octet S-wave LDMEs for a generic S-wave vector quarkonium state $V$ : $\Oosz,$ and $\Otsooct$. The leading transition operators that bring the color-octet $Q\bar{Q}$'s to a $^3S_1^{[1]}$ state were derived in ref.~\cite{Copeland:2025vop}.
By matching the LDMEs onto the transition operators and employing the Hubbard-Stratonovich transformation \cite{Fleming:2005pd}, we couple the quark and antiquark fields to the composite color singlet field. Applying this technique to the $\Oosz$ couples the operator to the color-singlet fields, at which point we can integrate out the quark and antiquarks to factorize the matrix element. This is shown in fig. \ref{fig: 1S08 quark loops}. 
The quark loop integrals were evaluated in Ref.~\cite{Copeland:2026yqa}, so we find the $\Oosz$ can be written as 
\begin{equation}
\begin{aligned}
   \Oosz = &\frac{1}{N_cM^2} \sum_{X_s} \bra{0}  {\cal S}_v^{\dagger,ab}\bigg[\frac{1}{v\cdot\Pc} g c_F {\cal B}^{i,b}_q(x)  \bigg][\sigma^i S_0^\dagger(x)]\ket{V ,X_s}\\
   &\times\bra{V ,X_s} [S_0(x)  \sigma^j]  {\cal S}_v^{ac}\bigg[\frac{1}{v\cdot\Pc} g c_F {\cal B}^{j,c}_q (x) \bigg]  \ket{0} \, .\\
\end{aligned}
\end{equation}
where $S_0$ is a color-singlet composite field, ${\cal S}_v^{\dagger,ab}$ is a v-directional soft Wilson line in the adjoint representation and ${\cal B}^{i,b}$ is a gauge invariant soft chromo-magnetic field. For reasons explained above, the soft Hilbert space separates from the composite field Hilbert space in production processes. Therefore, we can ``factorize" $\Oosz$ so that it takes the following form, 
\begin{equation}
\begin{aligned}
   \Oosz =
    &\frac{1}{3N_cM^2} |\bra{0} [\sigma^i S_0^\dagger(x)]\ket{V }|^2 \otimes \bra{0} \bigg|{\cal S}_v^{ab}\bigg[\frac{1}{v\cdot\Pc} g c_F {\cal B}^b_q  \bigg]\bigg|^2 (x) \ket{0} 
   , \\
\end{aligned}
\end{equation}
where we use spin and rotational symmetry to simplify the composite field matrix element. This factorization is represented by the right side of fig. \ref{fig: 1S08 quark loops}. Note, the matrix element with the chromo-magnetic field depends only on the soft scale and the matrix element with the composite color singlet field only depends on the ultrasoft scale. Defining the chromo-magnetic correlator $\braket{{\cal B}} $ \cite{Copeland:2026yqa}
we arrive at the factorized expression
\begin{equation}
\begin{aligned}
   \Oosz = &\frac{1}{3 N_cM^2} \braket{{\cal B}}\frac{3|R_V(0)|^2}{2\pi}   , \\
\end{aligned}
\label{eq: 1S08 fact}
\end{equation}
which agrees with the analysis from Ref.~\cite{Brambilla:2022ayc} exactly.

Likewise, matching the $\Otsooct$ onto the transition operators identified in Refs.~\cite{Copeland:2026yqa, Copeland:2025vop}, integrating out the quark and antiquark fields, and separating the Hilbert spaces, we find the result
\begin{figure}[t]
\centering
\vspace{-.7cm}
\includegraphics[width = .8\linewidth]{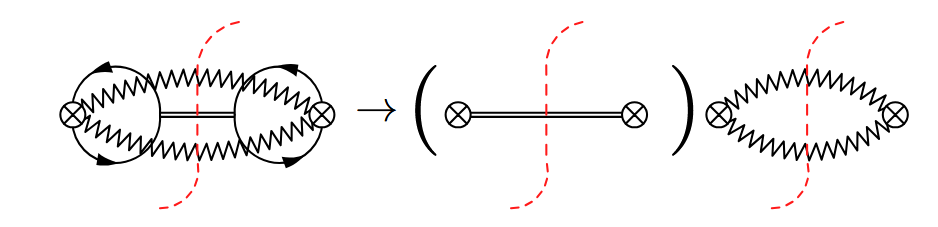}
\vspace{-.3cm}
    \caption{Factorizing out the soft double-electric dipole contributions in the $^3S_1^{[8]}$ channel. Zigzag lines represent the chromo-electric fields. Double lines represent the color-singlet composite field.}
    \vspace{-.5cm}
    \label{fig: 3S18 quark loops}
\end{figure}
\begin{equation}
\begin{aligned}
    \Otsooct =& \frac{1}{4N_cM^2} |\bra{0} [\sigma^i S_0^\dagger(x)]\ket{V }|^2 \otimes \bra{0} \bigg[ {\cal S}_v^{ag}\frac{1}{v\cdot\Pc}  \bigg(d^{gbc} g^2{\cal E}^{ b} \cdot {\cal E}^{c} \bigg)\bigg]^2\ket{0}.
\end{aligned}
\end{equation}
Defining the double electric field correlator $\braket{{\cal{EE}}}$ \cite{Copeland:2026yqa} we can write this in the compact form,
\begin{equation}
    \Otsooct = \frac{1}{N_cM^2} \braket{{\cal{EE}}}  \frac{3|R_V(0)|^2}{2\pi}
\label{eq: 3S18 fact}
\end{equation}
again reproducing the results from Ref.~\cite{Brambilla:2022ayc}.

In Ref.~\cite{Copeland:2026yqa}, it was found that the $\Otpz$ receives multiple operator contributions, each at a different order in the $\vb$ power counting. Matching the matrix element onto the leading vNRQCD operator, integrating out the quark loops, and separating the Hilbert spaces gives
\begin{equation}
\begin{aligned}
    \Otpz = & \frac{g^2}{12N_c } {\cal T}^{ij,i'j'}_J|\bra{0} [\sigma^i S_0^\dagger(x)]\ket{V }|^2 \otimes\bra{0} \big[ {\cal S}_v^{ab} {\cal E}^{j,b}\big]\big[ {\cal S}_v^{ab'}{\cal E}^{j,b'}\big]\ket{0}, 
\end{aligned}
 \end{equation}
where we have averaged over the Cartesian indices in the second line. 
where the ${\cal T}^{ij,i'j'}$ projectors are defined as \cite{Copeland:2026yqa}
\begin{equation}
\begin{aligned}
    {\cal T}_0^{ij,i'j'} &= \frac13 \delta^{ij}\delta^{i'j'} , \hspace{2cm}
    {\cal T}_1^{ij,i'j'} = \frac12 \epsilon^{kim}\epsilon_{ki'n}\delta^{mj}\delta^{nj'} , \\
    {\cal T}_2^{ij,i'j'} & = \bigg(\frac{\delta_{im}\delta_{nj}+ \delta_{in}\delta_{jm}}{3}-\frac13 \delta_{mn}\delta_{ij}\bigg)\bigg(\frac{\delta_{i'm}\delta_{nj'}+ \delta_{i'n}\delta_{j'm}}{3}-\frac13 \delta_{mn}\delta_{i'j'}\bigg). 
\end{aligned}
\label{eq: Pwave projectors}
\end{equation}

Finally, after averaging over spins and summing over polarizations, we can write the factorized P-wave matrix element as
\begin{equation}
    \Otpz = \frac{1}{36 N_c} (2J+1) \frac{3\big|R_V(0)\big|^2}{2\pi} \braket{{\cal E}} , 
\label{eq: leading 3PJ fact}
\end{equation}
where $ \braket{{\cal E}}$ is the chromo-electric field correlator \cite{Copeland:2026yqa}.
\begin{figure}[t!]
\centering
\vspace{-.7cm}
\includegraphics[width = .8\linewidth]{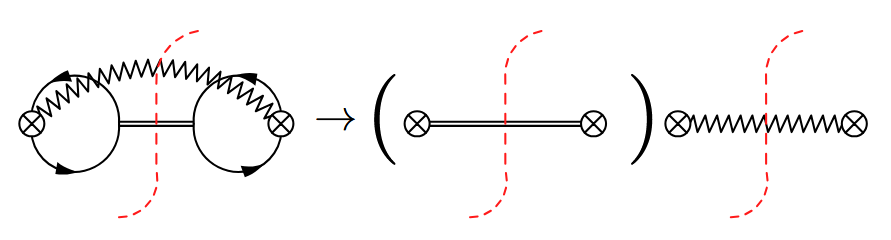}
\vspace{-.3cm}
    \caption{Factorizing out the soft chromo-magnetic dipole contributions in the $^3P_J^{[8]}$ channel. Zigzag line represents the chromo-electric field. Double lines represent the color-singlet composite field.}
    \vspace{-.5cm}
    \label{fig: 3PJ8 loops}
\end{figure}
%
%
Equation (\ref{eq: leading 3PJ fact}) agrees with the result determined using pNRQCD in ref.~\cite{Brambilla:2022ayc}. 

Next, we consider the subleading operator contribution identified in Ref.~\cite{Copeland:2026yqa}, which is a genuine P-wave operator at the soft scale and find instead 
\begin{equation}
    \Otpz^{(1)} = \frac{(m \braket{\vb^2}_V)^2}{108 N_c} (2J+1) \frac{3\big|R_V(0)\big|^2}{2\pi} \braket{{\cal E}^{(1)}} ,
\label{eq: gen 3PJ fact}
\end{equation}
where the result is now proportional to the binding energy of the quarkonium state ($m \braket{\vb^2}_V = M_V - 2m$) and a different electric field correlator, $\braket{{\cal E}}^{(1)}$ \cite{Copeland:2026yqa}.

In the TMD factorization framework, as the $p_T$ of the hadron approaches the soft scale of vNRQCD, the LDMEs should be replaced with $p_T$ dependent quarkonium production operators, known as TMDShFS and TMDSTFs. One problem with both the TMDShFs and TMDSTFs is that they encode soft radiation from other parts of the collision which cannot be separated from the soft radiation coming from the $Q\bar{Q}$. Therefore, these matrix elements have some inherent process dependence which spoils universality. This problem only arises if the hard scale of the collision $Q$ is similar to the mass of the quarkonium and the $P_T$ of the quarkonium is on the order of the soft scale of NRQCD. If $P_T \sim 2m \ll Q$, then the TMD production operators are matched onto the usual LDMEs \cite{Copeland:2023wbu, Copeland:2023qed, Echevarria:2023dme} and universality holds. 

Invoking the Hubbard-Stratonovich transformation, we can integrate out the quarks/antiquarks and write the leading TMDSTF as
\begin{equation}
\begin{aligned}
    &T^V_{^1S_0^{[8]} }(\bv_T;\mu,\eta)=\\
      &\qquad  \frac{1  }{M^2N_c\sqrt{S(\bv_T)}}\bra{0} \bigg[{\cal S}_v^{\dagger, bc} {\cal S}_n^{\dagger ba}  \frac{1}{v\cdot\Pc}   gc_F{\boldsymbol{\sigma} \cdot {\cal B}}_s^c \bigg]^i ({\bf b}_T)  \bigg[{\cal S}_n^{ b'a} {\cal S}_v^{b'c'}\frac{1}{v\cdot\Pc}   gc_F{\boldsymbol{\sigma} \cdot {\cal B}}_s^{c'} \bigg]^j(0) \ket{0}\\
      &\qquad \qquad  \times\bra{0}\big[\sigma^i S^\dagger_0(\bv_T)\big]\ket{V}\bra{V}\big[\sigma^j S_0(0)\big]\ket{0}.
\end{aligned}
\end{equation}
Since the composite field only carries ultrasoft momentum we can multi-pole expand away it's dependence on $\bv_T$.
After averaging over spins and summing over polarizations, we can write the TMDSTF in a simplified form,
\begin{equation}
    T^V_{^1S_0^{[8]} }(\bv_T; \mu, \eta) = \frac{1}{3N_cM^2} \braket{{{\cal B}_n}(\bv_T; \mu, \eta)} \frac{3|R_V(0)|^2}{2\pi} ,
\label{eq: fact TMDSTF}
\end{equation}
where $ \braket{{{\cal B}_n}(\bv_T;\mu, \eta)}$
is the process-dependent (but state-independent!) chromo-magnetic vacuum correlator \cite{Copeland:2026yqa}. This correlator can in principle be calculated on the lattice.

\section{Relations between different quarkonium LDMEs and applications to the TMD framework}

One powerful feature of these results is that the state-dependence of the LDME is moved into the wave function at the origin of the quarkonium, which is a well constrained quantity. The non-perturbative transition information is moved into the soft gluon correlators, $\braket{\cal B}, \braket{\cal E},$ and $\braket{\cal EE}$, which are universal state-independent quantities. So taking ratios of the expressions found in Eqs.~(\ref{eq: 1S08 fact}), (\ref{eq: 3S18 fact}), (\ref{eq: leading 3PJ fact}), and (\ref{eq: gen 3PJ fact}) for different S-wave vector quarkonium states, $V,V' = J/\psi, \psi(2S), \Upsilon(nS)$, cancels the correlator dependence and produces simple relations between the LDMEs of different states. Defining ${\cal R}_{VV'}^2 = |R_V (0)|^2/|R_{V'}(0)|^2$, we find
\begin{equation}
\begin{aligned}
    &\braket{O^V(^3S_1^{[8]})} = \frac{m_{Q'}^2}{m_{Q}^2} {\cal R}_{VV'}^2 \braket{O^{V'}(^3S_1^{[8]})} , \hspace{.5cm} \braket{O^V(^1S_0^{[8]})}= \frac{m_{Q'}^2}{m_{Q}^2}\frac{c_F^2(m_{Q})}{c_F^2(m_{Q'})}{\cal R}_{VV'}^2 \braket{O^{V'}(^1S_0^{[8]})},\\
    &\braket{O^V(^3P_J^{[8]})}= {\cal R}_{VV'}^2\braket{O^{V'}(^3P_J^{[8]})} 
    , \hspace{.5cm} \braket{O^V(^3P_J^{[8]})}^{(1)}= \bigg(\frac{m_{Q} \braket{\vb^2}_V}{m_{Q'} \braket{\vb^2}_{V'}}\bigg)^2{\cal R}_{VV'}^2 \braket{O^{V'}(^3P_J^{[8]})}^{(1)}.
\end{aligned}
\label{eq: LDME relations}
\end{equation}
where the scale dependence has been suppressed for brevity but can be found in Refs.~\cite{Copeland:2026yqa, Brambilla:2022ayc}. The first three of these relations were originally identified in Ref.~\cite{Brambilla:2022ayc}. The last relationship is for a novel {\it genunine P-wave} contribution at the soft scale \cite{Copeland:2026yqa} and may have important applications to hadroproduction of quarkonium at small $p_T$, since in this regime the P-wave contribution dramatically over predicts the data \cite{Brambilla:2024iqg}.

We find that the Hubbard-Stratonovich transformation techniques developed here can be used to derive similar relations between the TMDSTFs. For different states $V,V' = J/\psi, \psi(2S), \Upsilon(nS)$ we find
\begin{equation}
    T^V_{^1S_0^{[8]} }(\bv_T;\mu, \eta)  = \frac{m_{Q'}^2}{m_Q^2}\frac{c_F^2(m_Q,\mu)}{c_F^2(m_{Q'},\mu)}{\cal R}_{VV'}^2 T^{V'}_{^1S_0^{[8]} }(\bv_T;\mu, \eta).
\label{eq: TMDSTF relationship}
\end{equation}
This result is critical for the predictive power of NRQCD in the TMD framework. Since the TMDSTFs (and also the TMDShFs) are process-dependent quantities, they have almost no predictive power on their own. The TMDSTF is a non-perturbative quantity, particularly at low $p_T$, and the process dependence requires a unique TMDSTF for each new quarkonium TMD factorization theorem. However, the relationship in eq. (\ref{eq: TMDSTF relationship}) shows that the TMDSTFs for different quarkonium states can be related to each other through a re-scaling factor. For example, if $\Upsilon(1S)$ production at small $p_T$ was measured at the EIC, then in principle the TMDSTFs for $J/\psi$ produced at the EIC are well constrained through eq. (\ref{eq: TMDSTF relationship}). Therefore the only free parameter in a TMD factorization of the $e+p\to J/\psi +X$ cross section, would be the gluon TMDPDF - thereby enabling a clean extraction of this quantity.

\section{Conclusion}
In these proceedings, we have summarized powerful new results that have factorized the quarkonium production matrix elements in the collinear and TMD frameworks, the LDMEs and TMDSTFs. The Hubbard-Stratonovich transformations employed here show that the LDMEs and TMDSTFs can be written in terms of the wave function at the origin for the quarkonium state and universal chromo-electric and chromo-magnetic field correlators. We find that the leading order results for the LDMEs agree with results from the pNRQCD formalism in Ref.~\cite{Brambilla:2022ayc}. Moreover, we found novel sub-leading contributions to the P-wave matrix elements that were not included in the analysis of Ref.~\cite{Brambilla:2022ayc} and the TMDSTF results are completely new. These calculations permit powerful constraints that show that the LDMEs and TMDSTFs of different vector S-wave quarkonium states are related by a simple rescaling. This work reduces the amount of nonperturbative parameters in (v)NRQCD and dramatically improves the predictive power of the theory.

\vspace{.3cm}

{\bf Acknowledgments  } 
M.C. is supported by the U.S. Department of Energy through the Topical Collaboration in Nuclear Theory on Heavy-Flavor Theory (HEFTY) for QCD Matter under award no. DESC0023547. The work of M.C. is also supported by the US Department of Energy through the Los Alamos National Laboratory. Los Alamos National Laboratory is operated by Triad National Security, LLC, for the National Nuclear Security Administration of the U.S. Department of Energy (Contract No. 89233218CNA000001). Research presented in this article was supported by the Laboratory Directed Research and Development program of Los Alamos National Laboratory under project numbers 20260377ER and 20240131ER.

\end{document}